\documentclass[iop,apj]{emulateapj}

\newcommand{\Ms}{\ensuremath{M_{\odot}}}
\newcommand{\eg}{{\it e.g.}}
\newcommand{\cf}{{\it c.f.}}
\newcommand{\ie}{{\it i.e.}}
\newcommand{\viz}{{\it viz.}}

\slugcomment{In press with The Astrophysical Journal}

\shorttitle{NGC 3603 young cluster}
\shortauthors{Banerjee \& Kroupa}

\begin{document}

\title{A perfect starburst cluster made in one go: the NGC 3603 young cluster}

\author{Sambaran Banerjee and Pavel Kroupa}
\affil{Argelander-Institut f\"ur Astronomie, Auf dem H\"ugel 71, D-53121, Bonn, Germany;\\
sambaran@astro.uni-bonn.de, pavel@astro.uni-bonn.de}


\begin{abstract}
Understanding how distinct, near-spherical gas-free clusters of very young, massive stars shape out
of vast, complex clouds of molecular hydrogen is one of the biggest challenges in
astrophysics. A popular thought dictates that a single gas cloud fragments
into many new-born stars which, in turn, energize and rapidly expel the residual gas to form a
gas-free cluster. This study demonstrates that the above classical paradigm
remarkably reproduces the well-observed central, young cluster (HD 97950) of the Galactic NGC 3603
star-forming region, in particular,
its shape, internal motion and the mass distribution of stars, naturally and consistently follow from a single model
calculation. Remarkably, the same parameters (star formation efficiency, gas expulsion time scale and
delay) reproduce HD 97950 as were found to reproduce the Orion Nebula Cluster, Pleiades and R136.
The present results thereby provide intriguing evidences of formation of star clusters
through single-starburst events followed by significant residual gas expulsion.
\end{abstract}


\section{Introduction}\label{intro}

Very young, massive star clusters (hereafter VYMCs), which are compact associations of stars of mass $\gtrsim10^4\Ms$
and up to a few Myr old\footnote{Such clusters are usually referred to as ``starburst clusters'' but here we prefer
a more generic and origin-independent designation of such systems. They constitute a sub-category
of Young Massive Star Clusters \citep[]{pz2010}.},
are typically found in locations of high star-formation activity (or ``starburst regions'')
in our Milky Way and external galaxies. The classical notion implies that such systems form through
a single episode of starburst. In this scenario, efficient cooling processes within a dense parent gas-cloud
lead to the formation of a number of proto-stellar cores. Such an infant star cluster consists of massive main sequence (MS)
and lower-mass pre-main-sequence
(PMS) stars embedded within the residual gas. The radiation and
winds from the MS and PMS stars inject energy into this gas until the latter becomes unbound
and escapes the system.
The associated rapid dilution of the potential well causes the system to expand which
loses a fraction of its stars depending on its initial mass and concentration 
\citep[]{adm2000,pketl2001,bokr2002,bk2007,sb2013,pfkz2013}.
The remaining gas-free system  may eventually attain a state of virial equilibrium.
Such a ``monolithic'' formation scenario has
successfully explained the structure and kinematics
of Galactic and extra-galactic VYMCs like the Orion Nebula Cluster (ONC) and R136
and intermediate-aged clusters like the Pleiades and Hyades \citep[]{pketl2001,sb2013}.

In this study we incorporate the same properties and parameters (see below)
of the gas expulsion process as those in the earlier studies \citep[]{pketl2001,sb2013}
in our realistically detailed model calculations of
the time-evolution of star clusters. We find good and consistent agreement between
these model computations and detailed measurements of the central young cluster (HD 97950; hereafter HD97950)
of the Galactic starburst region NGC 3603 which are based on infrared and optical observations with the ESO/VLT
(Very Large Telescope) and the Hubble Space Telescope (HST), respectively \citep[]{hara2008,roch2010,pang2013}.
In particular, we
closely reproduce its observed (a) mass density profile \citep[]{hara2008},
(b) central velocity dispersion \citep[]{roch2010,pang2013},
(c) present-day stellar mass function (PDMF; \citealt[]{pang2013}), and as well (d) the spatial distribution
of stars from the crude data \citep[]{pang2013}, simultaneously from one individual model calculation.
This not only strongly points to a single-episode formation of HD97950 as envisaged previously \citep[]{stol2004,stol2006}
but it also affirms the monolithic formation channel for VYMCs in general and suggests that such systems
evolve according to universal principles.

\section{Model calculations}\label{compute}

\subsection{Initial systems and gas dispersal}\label{initsys}

We perform time ($t$)-evolutionary calculations that mimic gas-dispersal
from model star clusters beginning from their
gas-embedded phase. The star clusters are initialized by distributing stars over a highly compact (see below)
and spherically-symmetric Plummer phase-space distribution \citep[]{hh2003,pk2008} of total stellar mass $M_{ecl}$.
The stellar initial mass function (IMF) follows the canonical two-part power-law \citep[]{pk2001,pk2013}.
The dominant physical effect of the embedding gas is taken into account by applying an external
potential from a spherically-symmetric
mass distribution of total mass $M_g$ following the same Plummer distribution as the stars.
The gas-depletion is implemented by asymptotically reducing $M_g$ with a timescale $\tau_g$ after
an embedded-phase time-span of $\tau_d$ as in \citet[]{pketl2001,sb2013}.
Detailed \emph{Herschel} observations
indeed indicate Plummer-like sections \citep[]{mali2012} of highly compact (about 0.1 pc in section)
gas filaments within which (and at their intersections) the spherical/spheroidal proto clusters form
via gravitational fragmentation and collapse \citep[]{schn2010,schn2012,hill2011,henne2012}.
Adopting an appropriately time-varying background potential, mimicking the effect of the gas,
is currently the only viable method
to computationally deal with the complex magneto-hydrodynamic plus radiation transfer
problem on this mass scale (see Sec.~\ref{discuss}).

\subsubsection{Gas dispersal}\label{gdisp}

\begin{figure}
\includegraphics[width=9.2cm, angle=0]{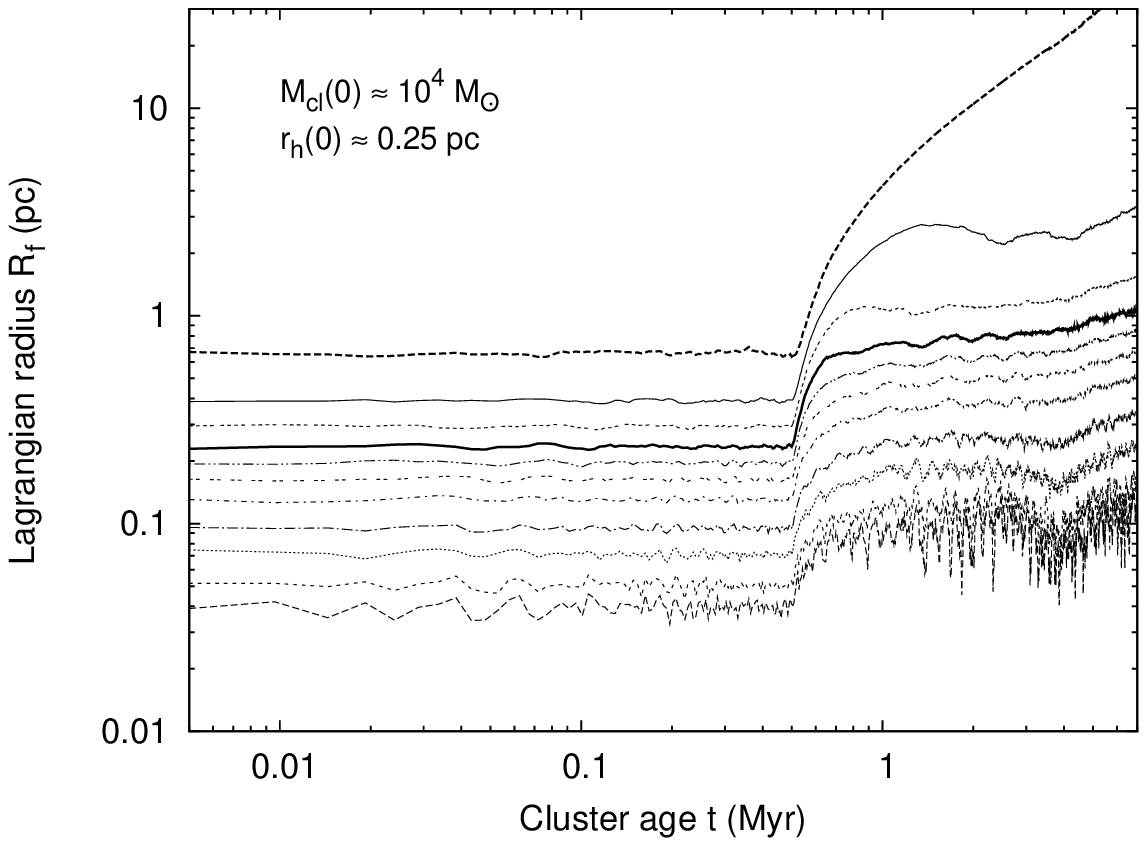}
\includegraphics[width=9.2cm, angle=0]{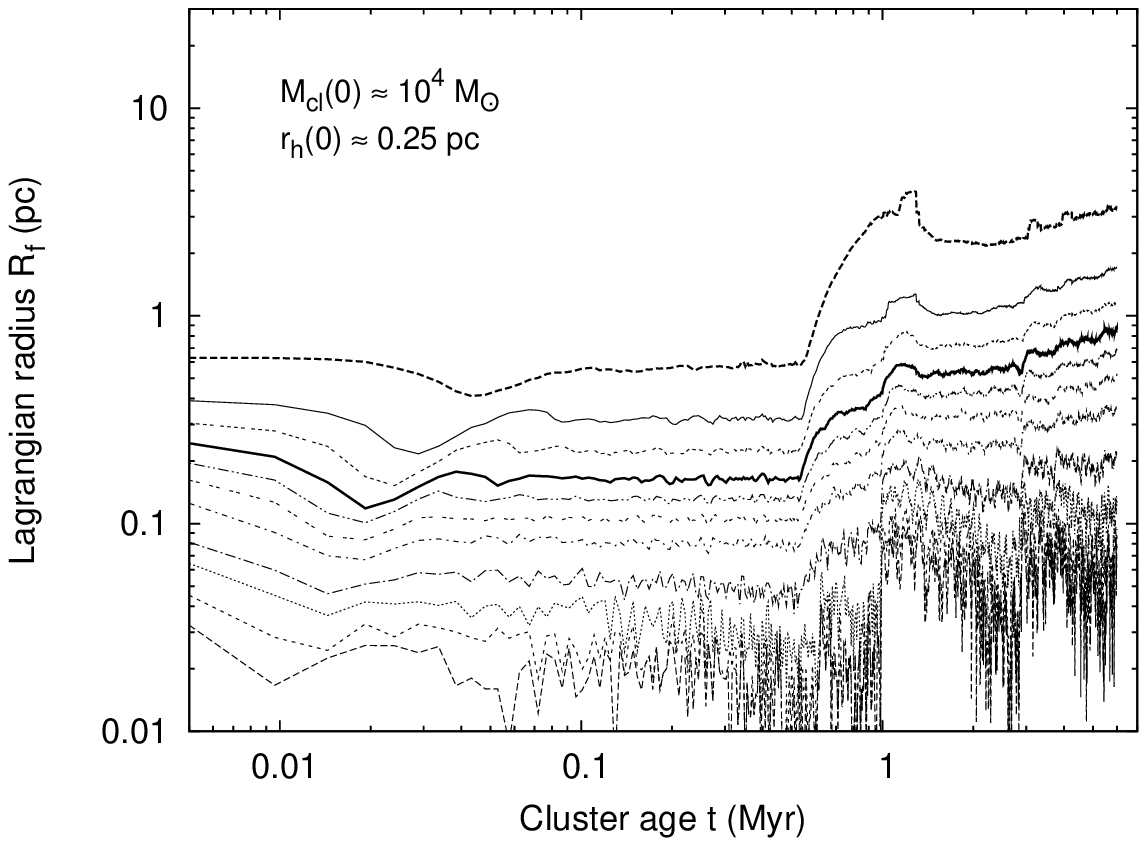}
\caption{Lagrange radii $R_f$ for mass fraction $f$ 
for the model cluster computations without (top; model HD97950s in Table~\ref{tab1})
and with (bottom; model HD97950b in Table~\ref{tab1}) primordial binaries (Sec.~\ref{compute}).
In each panel, the curves, from bottom to top, correspond to $f=0.01,0.02,0.05,0.1,0.2,0.3,0.4,0.5,0.625,0.7,0.9$
respectively. The Lagrange radii evolution shows a rapid expansion of the clusters at $t_d\approx0.6$ Myr
due to gas expulsion followed by re-collapse of a fraction of them to dynamical equilibrium.
For the computation with binaries (Sec.~\ref{primbin}), the cluster undergoes a brief initial collapse phase
or ``binary cooling'' (Sec.~\ref{bincomp}). In the top panel (no primordial binaries), the $R_{0.9}$
blows up due to the inclusion of escapers.}
\label{fig:lagrcomp}
\end{figure}

\begin{deluxetable*}{lccccccccl}
\tablewidth{7 in}
\tablenum{1}
\tablecaption{Initial and gas expulsion parameters for the computed
models that reproduce well observed VYMCs (very young massive clusters;
see Sec.~\ref{intro}).
The models HD97950s/b refer to the ones with initially single-only stars/primordial binaries as computed
here (see Sec.~\ref{res}).
Also included are the previously published model clusters that agree well with the VYMCs R136 \citep[]{sb2013}
and ONC (and also Pleiades; \citealt[]{pketl2001}). 
}
\label{tab1}
\tablecolumns{10}
\tablehead{
\colhead{Model cluster} & \colhead{$M_{\rm ecl}(0)/\Ms$} & \colhead{$M_g(0)/\Ms$} & \colhead{$r_h(0)/{\rm pc}$}
 & \colhead{$\tau_g/{\rm Myr}$} & \colhead{$\tau_{\rm cr}(0)/{\rm Myr}$} & \colhead{$\tau_d/{\rm Myr}$}
 & \colhead{$f_b(0)$} & \colhead{$Z/Z_\odot$} & \colhead{Reference}
}
\startdata
HD97950s & $1.0\times10^4$ & $2.0\times10^4$ & 0.25 &  0.025 & 0.029  & $0.6$ & 0.0 & 1.0 &this paper\\ 
HD97950b & $1.0\times10^4$ & $2.0\times10^4$ & 0.25 &  0.025 & 0.025  & $0.6$ & 1.0 & 1.0 &this paper\\ 
\tableline                                                                               
R136   & $1.0\times10^5$   & $2.0\times10^5$ & 0.45 &  0.045  & 0.021 & 0.6 & 0.0 & 0.5 &\citet[]{sb2013}\\
ONC-B  & $4.2\times10^3$   & $8.4\times10^3$ & 0.21 &  0.021  & 0.066 & 0.6 & 1.0 & 1.0 &\citet[]{pketl2001}
\enddata
\tablecomments{The initial gas mass $M_g(0)$ and the gas expulsion timescales
$\tau_g$ and $\tau_d$ (see Sec.~\ref{gdisp}) are consistent with $\epsilon\approx0.33$, $v_g\approx10{\rm~km~s}^{-1}$
($\tau_g=r_h(0)/v_g$) and $\tau_d\approx0.6$ Myr for all these computed models. See Sec.~\ref{discuss} for further discussions.}
\end{deluxetable*}

We adopt the same values of the above parameters quantifying the overall gas expulsion as in
\citet[]{pketl2001} and \citet[]{sb2013}.
Both observations \citep[]{lnl2003} and theoretical studies \citep[]{mnm2012} support a local star-formation efficiency (SFE) of
$\epsilon\approx1/3$, \ie, $M_g(0)=2M_{ecl}(0)$ initially ($t=0$; see more in Sec.~\ref{discuss}).

The timescales $\tau_d$ and $\tau_g$
depend on the complicated physics of gas-radiation interaction.
For simplicity,
we estimate $\tau_g$ by taking the average radial speed
of the expelling gas to be the sound speed in ionized hydrogen (\ie, that of an HII region),
\viz, $v_g\approx10{\rm~km~s}^{-1}$. This sound speed corresponds to the typical temperature
of $\approx10^4$ K of an HII region \citep[]{ostb1965,andrson2009}\footnote{
$T\approx10^4$ K is the thermal equilibrium temperature where the heating by the ionizing
UV radiation (primarily from the OB stars) is compensated by the cooling due to the free electrons
(mostly via free-free and free-bound transitions) in the ionized medium. Neutral hydrogen (HI) regions,
on the contrary, are much colder ($<100$ K). The effective ionization in HII (by UV) causes efficient
coupling of the radiation flux with the free electrons and hence results in much more efficient energy transfer
than that in HI.}.
In other words, $\tau_g=r_h(0)/v_g$, where $r_h(0)$ is the
initial half-mass radius ($\approx$ virial radius for a Plummer model) of the stellar
cluster/gas. The residual gas begins to launch when it becomes ionized by the
(proto-)stellar radiation so that the latter is coupled efficiently with the gas.
The coupling of the radiation with the resulting HII medium substantially overpressures the latter
and can even make it radiation-pressure-dominated (RPD) which might briefly drive the
gas supersonically in the beginning \citep[]{krm2009}. The expanding gas continues to move out with the sound speed
of an HII region once it becomes gas-pressure-dominated (GPD) \citep[]{hill80}. This initial
RPD phase is essential for launching the gas from massive clusters whose escape speed exceeds the HII sound speed. 
Therefore, for massive clusters, the above $\tau_g$, from $v_g\approx10$ km s$^{-1}$, gives an upper limit
(also see \citealt[]{pk2005}); it can be shorter depending on the duration of the RPD state (also see Sec.~\ref{discuss}).

This $\tau_g$ is typically comparable to the dynamical time (or crossing time over half-mass radius) $\tau_{\rm cr}$ of the
embedded cluster. This makes the gas expulsion process ``explosive'' \citep[]{pk2005}, where the cluster is unable
to adjust itself to the varying gas potential and hence expands ``violently''. It can be shown \citep[]{pk2005} that for
moderate to massive clusters,
the energy (wind+radiation) injected in the gas by the OB stars within a dynamical time
is sufficient to make the gas unbound, supporting an explosive nebula disruption. 

As for the delay-time, we take $\tau_d\approx0.6$ Myr as a representative value
\citep[]{pketl2001}. The value of $\tau_d$ again depends on gas-radiation interaction. An estimate
of $\tau_d$ can be obtained from the lifetimes of the Ultra Compact HII (UCHII) regions which can be
upto $\approx 10^5$ yr (0.1 Myr; \citealt[]{chrch2002}). The highly compact pre-gas-expulsion clusters (Sec.~\ref{res})
have sizes ($r_h(0)$) only a factor of $\approx3-4$ larger than the typical size of a UCHII region ($\approx0.1$ pc).
If one applies a Str\"omgren sphere expansion scenario (\citealt[]{chrch2002} and references therein)
to the compact embedded cluster, the estimated delay-time, $\tau_d$, before a sphere of radius $r_h(0)$
becomes ionized, would also be larger by a similar factor and hence close
to the above representative value. Once the
gas gets ionized, it couples efficiently with the radiation (pressure) from the O/B stars and is
launched immediately (see above).
High-velocity outflows from proto-stars \citep[]{pat2005} additionally aid the gas expulsion. A more
elaborate calculation of the expansion of a Str\"omgren front through the proto-cluster gas
is underway (Pflamm-Altenburg, in preparation).

Admittedly, the above arguments ignore complications such as unusual morphologies
of UCHIIs and possibly non-spherical ionization front, among others, and only provide basic estimates
of the gas-removal timescales. Observationally,
Galactic $\approx 1$ Myr old gas-deprived VYMCs such as the ONC and the HD97950 imply that
the embedded phase can be $\tau_d<1$ Myr. The above widely-used analytic gas-expulsion model
does realize the essential dynamical effects
on the star cluster after the residual gas is depleted.
Moreover, the above same values of $\epsilon\approx0.33$, $\tau_d\approx0.6$ Myr and $\tau_g=r_h(0)/v_g$
($v_g\approx10{\rm~km~s}^{-1}$)
that reproduced the ONC and the Pleiades \citep[]{pketl2001}
as well as the R136 \citep[]{sb2013} also fair excellently
with the HD97950 as we shall see in the following sections.

\subsubsection{Primordial binaries}\label{primbin}

Apart from initial systems consisting of only single stars as above, we also create initial systems
with primordial binaries which is the realistic case. Here
we use a $f_b(0)=100$\% primordial binary fraction at $t=0$ \citep[]{pk1995a}.
For stars of mass $m<5\Ms$, the orbital period ($P$) distribution of such binaries
is given by the ``birth period distribution'' \citep[]{pk1995a}
which spans over a wide range of period between
$1.0 < \log_{10} P < 8.43$ where $P$ is in days \citep[]{pk1995b}.
We take a thermal distribution of binary eccentricities.
With the dynamical evolution and eventual
disruption (by the Galactic tidal field) of the parent cluster, such primordial binary population
gives rise to the appropriate period distribution observed for low-mass 
stellar binaries in the solar neighborhood \citep[]{pk1995a,mk2011}.

For massive stars of $m>5\Ms$, we adopt a much tighter and narrower initial period distribution
given by a (bi-modal) \"Opik law (uniform distribution in $\log_{10} P$) that
spans over the range $0.3 < \log_{10} P < 3.5$ \citep[]{sev2011}. This is motivated
by the observed period distribution of O-star binaries in nearby O-star rich clusters \citep[]{sev2011,chini2012}.
It is currently unclear at which stellar mass the orbital period law changes and how it changes and
the above discontinuous switching of the $P$-distribution at $m=5\Ms$, therefore, had to be chosen
somewhat arbitrarily. The consistency of the resulting break in the stellar mass function of
the computed clusters with the observed
one (see Sec.~\ref{bincomp}) provides a justification to the above switching. Apart from the
MF-break, our results, as discussed below,
are largely insensitive to the switching point of the period law (in the range $2\Ms-7\Ms$, see below).

For $m>5\Ms$, we furthermore use biased/ordered pairing in which the stars, picked
by optimally sampling \citep[]{pk2013} the canonical IMF, are arranged in order of $m$
and paired sequentially. This is motivated by the observation that O-stars
are typically found in binaries with O/B stars \citep[]{sev2011,chini2012}. For lower $m$, the stars are paired
by picking each component randomly and independently from the canonical IMF. 
This is consistent with all available data on young and old Galactic field systems \citep[]{mk2011}.
Finally, for very low-mass stellar binaries with short ($P<10^3$ d) periods,
the orbits are adjusted to account for PMS ``eigenevolution'' \citep[]{pk1995b}.
This takes into account the very early (age $<10^5$ yr) system-internal orbital energy
and orbital angular momentum redistribution due to, \eg, tidal circularization and
system-internal residual gas.

\subsection{N-body computations}\label{nbcomp}

We trace the time evolution of such initial systems (set using {\tt MCLUSTER}; \citealt[]{kup2011}) 
using the state-of-the-art direct N-body integration code {\tt NBODY6} \citep[]{ars2003}.
Apart from utilizing a highly sophisticated numerical
integration scheme, {\tt NBODY6} also keeps track of the evolution of the individual stars \citep[]{hur2000}.
We use the iron metallicity of $Z=Z_\odot$ in all the calculations, as appropriate for HD97950.
We explore a range of plausible initial
clusters to spot one that reproduces the observed structure and kinematics
of HD97950, being consistent with its age and photometrically-determined mass ranges
of 1-2 Myr and $10000\Ms-16000\Ms$, respectively \citep[]{stol2004,stol2006}.

\section{Results: a surprisingly well-matching model}\label{res}

\begin{figure}
\includegraphics[width=9.2cm, angle=0]{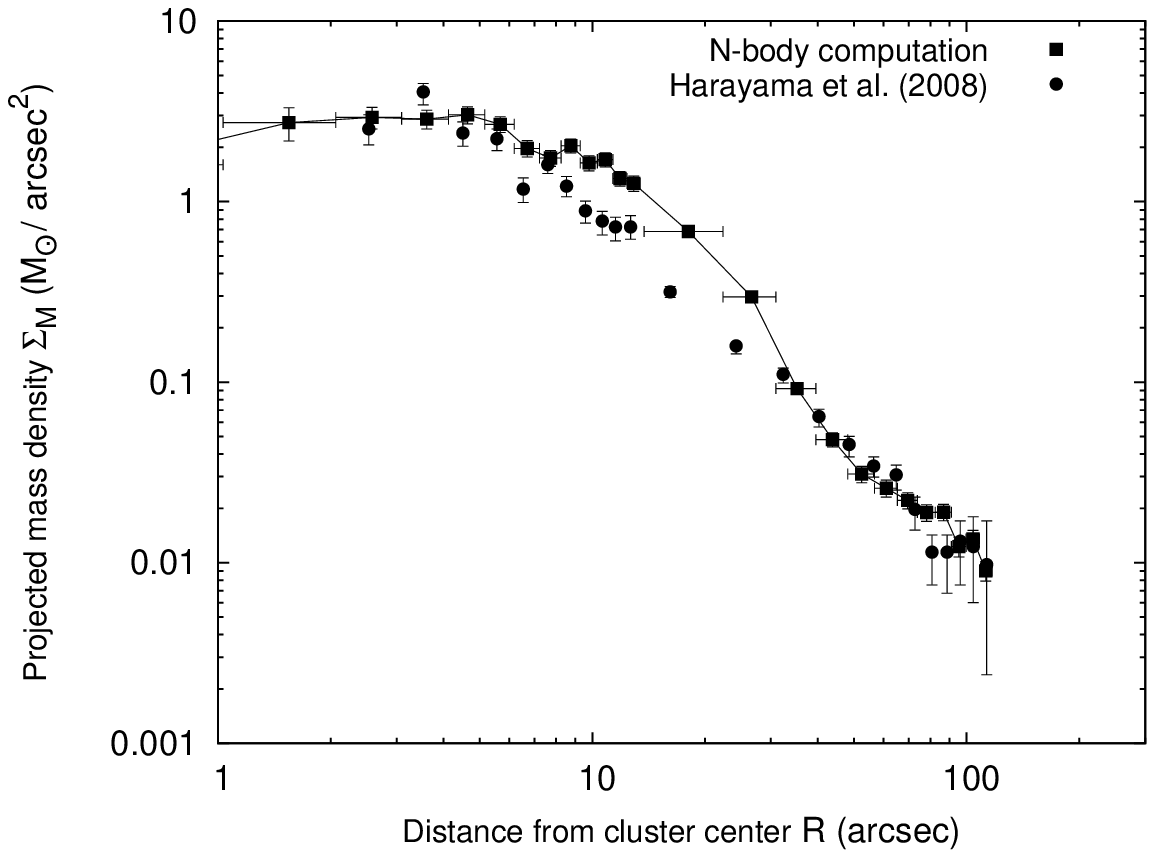}
\includegraphics[width=9.2cm, angle=0]{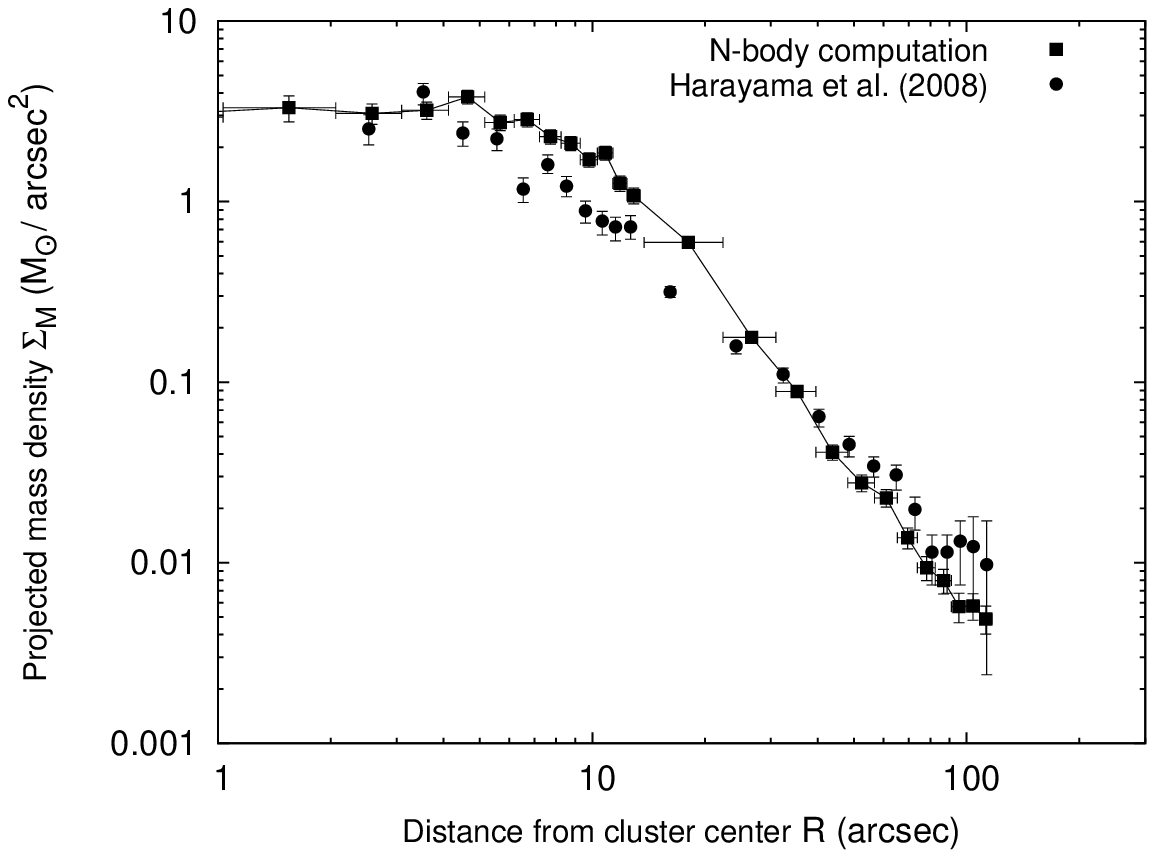}
\caption{{\bf Top:} The computed projected mass density profile (filled squares and solid line)
for the stellar mass range $0.5\Ms<m<2.5\Ms$ at $t\approx1.2$ Myr from a model computation
containing only single stars initially ($r_h(0)\approx0.25$ pc, $M_{ecl}(0)\approx10000\Ms$;
same model HD97950s as shown in the top panel of Fig.~\ref{fig:lagrcomp}).
This profile shows remarkable agreement with the observed profile \citep[]{hara2008},
for the same stellar mass range, of the central young cluster (HD97950)
of the NGC 3603 starburst region (filled circles). The angular annuli used for computing the projected
densities (the horizontal error-bars) are nearly the same as those used to obtain the observed profile.
{\bf Bottom:} The computed profile of a star cluster having the same mass and size as above initially but
with all stars in primordial binaries, showing
excellent agreement with the observed profile at $t\approx1.4$ Myr. All legends are the same as above.
This is the same model HD97950b as shown in the bottom panel of Fig.~\ref{fig:lagrcomp}.
Here, and for all the next figures, $1{\rm~arcsec}\approx0.03$ pc as appropriate for HD97950.}
\label{fig:Mdensprof}
\end{figure}

As expected, all computed clusters undergo a nearly steady, compact embedded state 
until the gas expulsion at $\tau_d\approx0.6$ Myr from which they begin to expand
(see Fig.~\ref{fig:lagrcomp}, top panel). The cluster
profile correspondingly changes from a highly concentrated and relatively static one
to an expanded and rapidly changing profile. Fig.~\ref{fig:Mdensprof} (top, solid line and filled squares)  
shows the surface/projected mass density profile $\Sigma_M$ at $t\approx1.2$ Myr for a computed cluster of only single
stars with half-mass radius $r_h(0)\approx0.25$ pc
and $M_{ecl}(0)\approx10000\Ms$ for the stellar component at $t=0$ as
shown for the model ``HD97950s'' in Table.~\ref{tab1} (hence $r_h(0)\approx0.25$ pc
and $M_g(0)\approx20000\Ms$ for the gas, $\tau_g\approx0.025$ Myr which is close to the initial
dynamical time $\tau_{\rm cr}(0)\approx0.029$ Myr).
It agrees remarkably with that observed in
HD97950 (filled circles; \citealt[]{hara2008}). Note that in this comparison a similar stellar mass range
and annuli as those for the observed profile are used to construct the density profile
from the computed cluster.

How well does this matching model reproduce other observed properties of HD97950? An important check is to
compare with the cluster's central stellar velocity dispersion which is obtained through
proper motion measurements \citep[]{roch2010,pang2013}.
The above computed model,
however, gives a rather lower (1-dimensional) velocity dispersions ($\sigma_{1d}\approx3{\rm~km~s}^{-1}$)
for $1<t<2$ Myr
than what is observed in HD97950, \viz, $\sigma_{1d}=4.5\pm0.5{\rm~km~s}^{-1}$ for stars of masses between $1.7\Ms<m<9.0\Ms$
within $R<15''(\approx0.5{\rm~pc})$ from the cluster center \citep[]{roch2010}. This is shown in Fig.~\ref{fig:v1d_pang} (top).

\subsection{Computations with primordial binaries}\label{bincomp}

\begin{figure}
\includegraphics[width=9.2cm, angle=0]{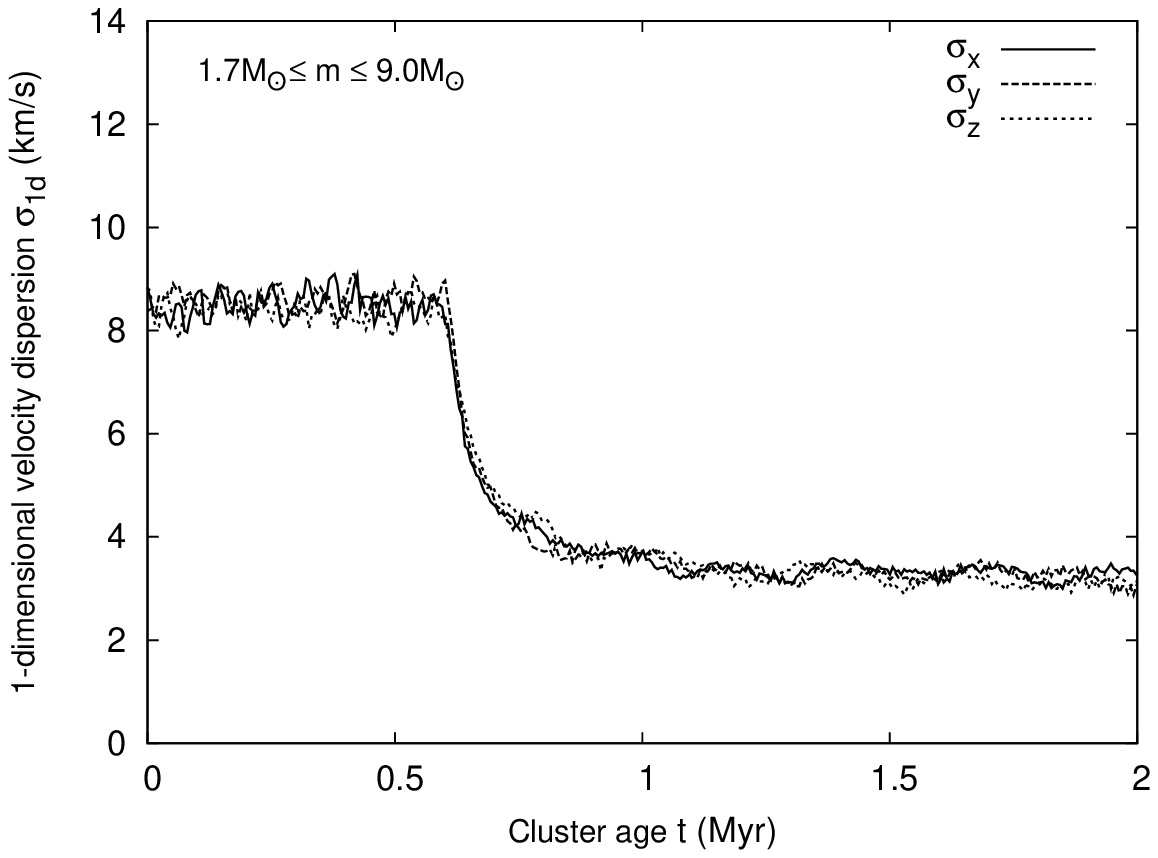}
\includegraphics[width=9.2cm, angle=0]{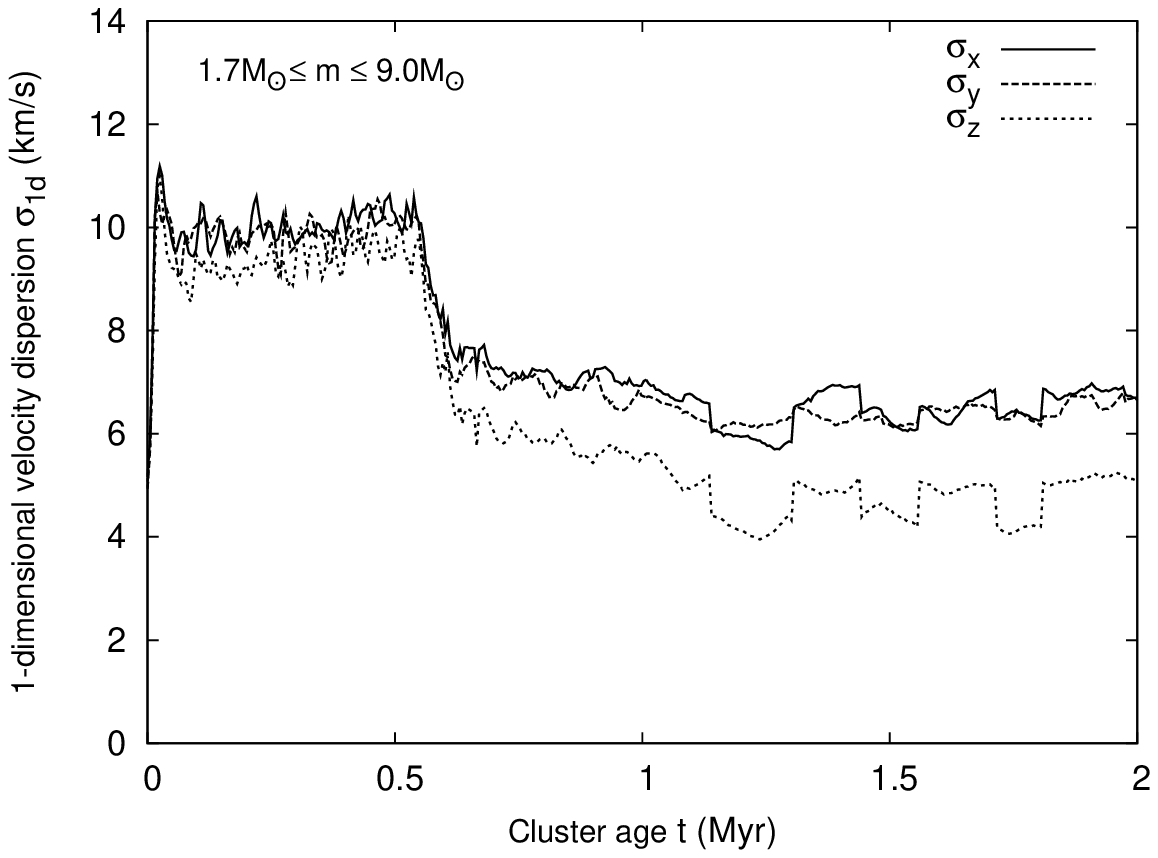}
\includegraphics[width=9.2cm, angle=0]{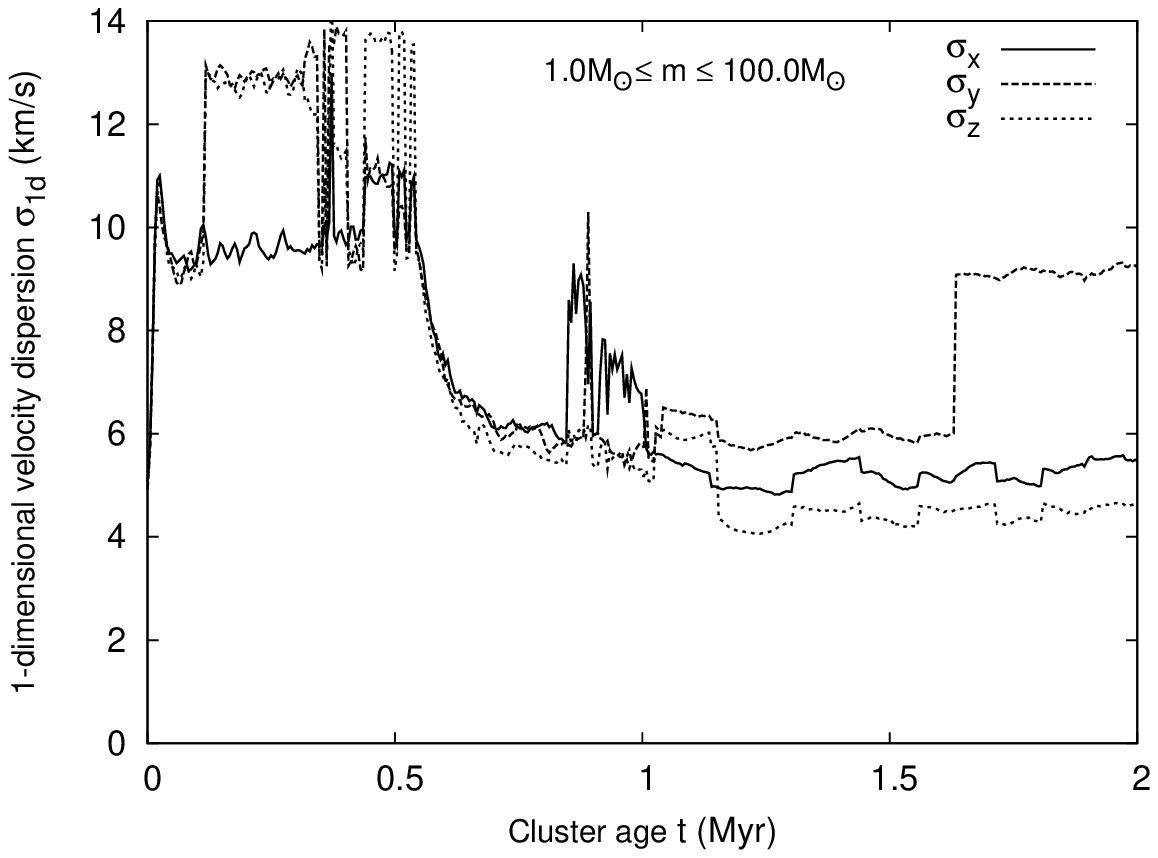}
\caption{
The time evolution of the 1-dimensional velocity dispersions, $\sigma_{1d}$,
for the model cluster computations without (top) and with (middle)
primordial binaries (see text), the density profiles of which are shown
in Fig.~\ref{fig:Mdensprof}. 
They are obtained for $R<0.5$ pc ($\approx15''$) and $1.7\Ms<m<9.0\Ms$
consistently with the observations of \citet[]{roch2010}.
The bottom panel shows the $\sigma_{1d}$ evolution in the latter computation but
for $1.0\Ms<m<100.0\Ms$ as in \citet[]{pang2013}. For the binaries in the computed model,
the center-of-mass velocities, giving their overall motion through the cluster, has been
used while their most massive/brightest member stars (primaries) refer to the masses. The
computed values of $\sigma_{1d}$ (bottom panel) differ in orthogonal directions
as found in observations \citep[]{pang2013} and span the same range as observed
(4.5-7.0 ${\rm~km~s}^{-1}$) between 1.0 - 1.5 Myr cluster age, implying good agreement.}
\label{fig:v1d_pang}
\end{figure}

Interestingly, this
deficiency is compensated if one instead considers the computed models with primordial binaries (Sec.~\ref{primbin}).
Fig.~\ref{fig:v1d_pang} (middle) shows the $\sigma_{1d}$ evolution for a computed model
with $f_b(0)=100$\% primordial binary fraction but otherwise identical to the above single-star model at $t=0$
(model ``HD97950b'' in Table~\ref{tab1}),
for the same zone and mass range as above. The augmented values of
$\sigma_{1d}$ (of the COMs, \ie, excluding the binaries' internal orbital motion) in this case
can be attributed to frequent binary-single and binary-binary encounters that release the internal
binding energy of the binaries (``binary-heating''; \citealt[]{hh2003}). Here,
the values of $\sigma_{1d}$ lie between $4.5-7.0{\rm~km~s}^{-1}$ which includes the observed value.
Notably, the models with primordial binaries undergo an initial contraction by
$\approx 10$\% as shown in Fig.~\ref{fig:lagrcomp} (bottom),
unlike the single-star models. This is caused by the kinetic energy absorbed in
ionizing loosely-bound or ``soft'' binaries which are present in large numbers
initially (\citealt[]{pk1995a}; Sec.~\ref{primbin}). This effect can therefore be referred to as
``binary-cooling'' \citep[]{pk1999}.
In fact, the presence of a large number of soft binaries effectively cause a moderately sub-virial
initial system which collapses and virializes (after re-expanding; \cf, Fig.~\ref{fig:lagrcomp})
in a few crossing times. The presence of a gas potential deepens this initial collapse.
Note that this effect is a direct result of the chosen initial binary
$P$-distribution (\cf, Sec.~\ref{primbin}). This
distribution is successful in reproducing the observed $P$-distribution of low-mass stellar 
field binaries in the solar neighborhood \citep[]{pk1995a,pk1995b,mk2012} and hence constitutes
a realistic initial condition. The details of this process will be studied in a future paper.

However, the corresponding initial collapse does not conflict with the intended $\epsilon\approx0.33$
as this SFE corresponds to that of the overall system, \ie, the total star-to-gas ratio of the whole initial cluster.
This is what determines the post-gas-expulsion violent relaxation (expansion) of the cluster. 
This initial shrinkage would cause the cluster to attain the observed density profile somewhat later in time (see below).

A recent work (Banerjee et al., in preparation), in which an HD79750-like massive cluster forms
via merging of much less-massive sub-clusters within a gas potential, shows that a similarly compact cluster
as in the present case is formed right before gas expulsion. 
The formation of such a massive cluster is, of course, true only if the implied gas density is very high.

\begin{figure}[!ht]
\includegraphics[width=9.2cm, angle=0]{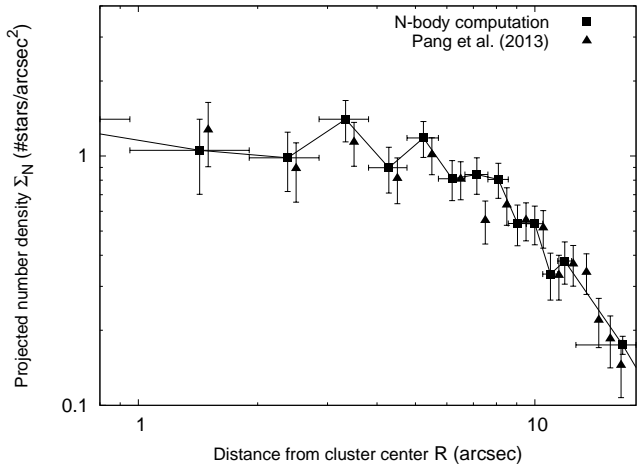}
\caption{The computed projected stellar number density profile for the calculation 
HD97950b (Table~\ref{tab1}) with
primordial binaries (filled squares and solid line; c.f. Fig.~\ref{fig:Mdensprof}, text). This
shows a remarkable agreement at $t\approx1.4$ Myr with the same obtained with the stars of
HD97950 from the HST/PC chip (extends
up to central $15''$; filled triangles; \citealt[]{pang2013}). In constructing the computed profile,
we take into account the incompleteness in the detection of the stars that depends on the
stellar mass (luminosity) and projected angular annuli on the cluster \citep[]{pang2013}. In this comparison,
similar angular annuli (horizontal error-bars) have been used to construct the density profiles.}
\label{fig:Ndensprof}
\end{figure}

The above computed model with binaries as well reproduces the observed projected mass density profile
at $t\approx1.4$ Myr, shown in Fig.~\ref{fig:Mdensprof} (bottom).
Fig.~\ref{fig:Ndensprof} shows the profile of the
incompleteness-limited stellar number density $\Sigma_N$ from the above computed cluster (filled squares and
solid line) at $t\approx1.4$ Myr and that obtained from HST \citep[]{pang2013} (filled triangles) which are in
excellent agreement. In this comparison, we have sampled the computed distribution
according to the radius and mass-dependent incompleteness fraction particular to this observation \citep[]{pang2013},
to mimic the ``observation'' of the model cluster. In constructing both of these density profiles
we include only the most massive member (primary) of a binary which would dominate
the detected light from it.

Fig.~\ref{fig:v1d_pang} (bottom) shows the $\sigma_{1d}$ evolution
in the binary model over $1.0\Ms<m<100.0\Ms$ within $R<0.5$ pc. The computed dispersion of the velocity
components lie between $4.0<\sigma_{1d}<7.0{\rm~km~s}^{-1}$ for $1<t<2$ Myr. The corresponding 
measured 1-d velocity dispersions do change substantially with orthogonal directions \citep[]{pang2013}
like the computed ones (see Fig.~\ref{fig:v1d_pang}; bottom panel) and their variations closely
agree to the above computed range. The abrupt vertical excursions in $\sigma_{1d}$ in Fig.~\ref{fig:v1d_pang} (bottom)
originate from energetic two- or few-body encounters which are most frequent for
the most massive stars and binaries as they centrally segregate the most via two-body relaxation.

\begin{figure}
\includegraphics[width=9.2cm, angle=0]{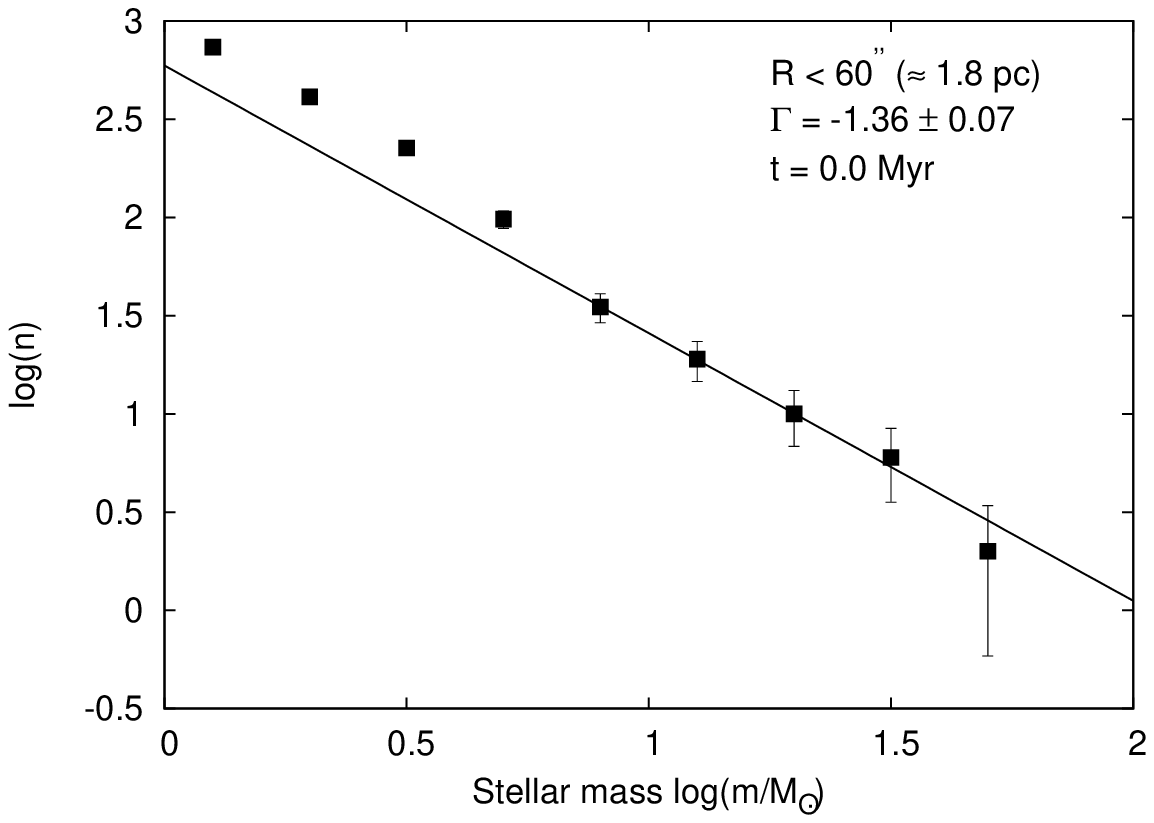}
\includegraphics[width=9.2cm, angle=0]{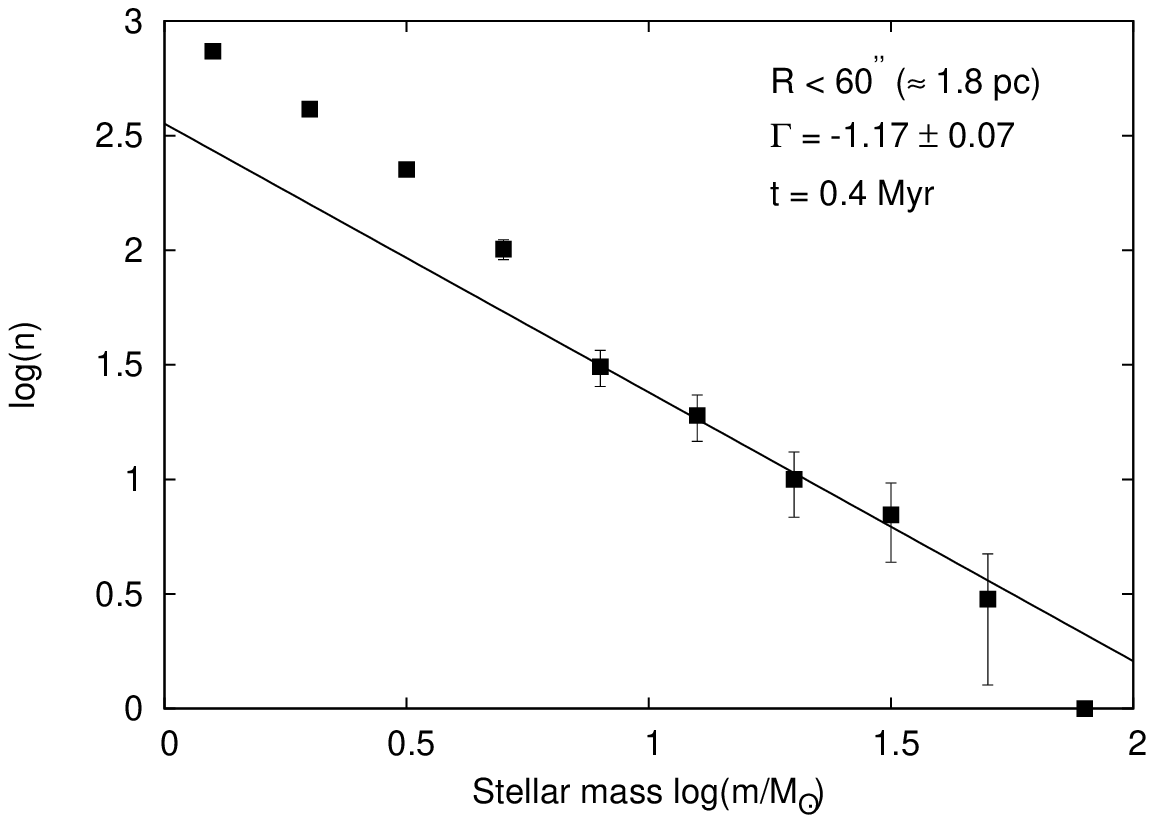}
\includegraphics[width=9.2cm, angle=0]{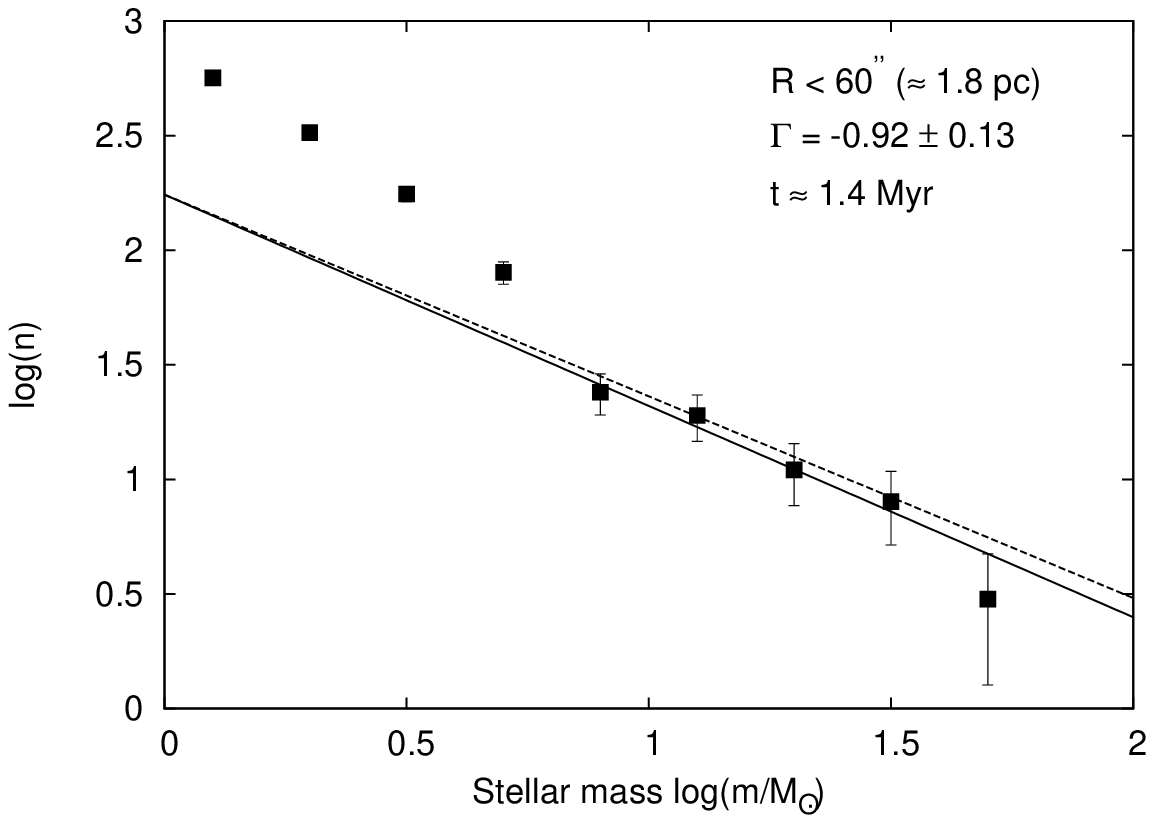}
\caption{The stellar mass function (of primaries)
within the central $60''$($\approx 1.8$ pc) of the computed model with primordial binaries (model HD97950b)
at $t\approx0.0$, 0.4 and 1.4 Myr (top, middle and bottom panels respectively).
Beginning with the global
canonical value of $\Gamma=-1.35$ ($\alpha_2=2.35$), the MF slope develops a break with time
as a result of the switching to much tighter initial orbital
periods for massive binaries (from $>5\Ms$; see Sec.~\ref{bincomp}).
Upwards the break (at $\log(m/\Ms)\approx0.7$),
the MF slope becomes increasingly shallow while it remains close to canonical for lower $m$.
The corresponding slope at $t\approx1.4$ Myr (bottom), \viz,
$\Gamma=-0.92\pm0.13$ (solid line), agrees closely with the corresponding
measured one, \viz, $\Gamma=-0.88\pm0.15$ (dashed line; \citealt[]{pang2013}).}
\label{fig:mfnow}
\end{figure}

Finally, Fig.~\ref{fig:mfnow} (bottom) shows the computed global ($R<60''$ or $\approx1.8$ pc) stellar mass function
(again, of only the primaries) or the present day mass function (PDMF) at $t\approx1.4$ Myr (filled squares). 
The PDMF shows a vivid break at $\log(m/\Ms)\approx0.7$ beyond which it becomes much shallower,
with slope $\Gamma=-0.92\pm0.13$, than the canonical IMF. This $\Gamma$ is in marked agreement with 
the observed $\Gamma=-0.88\pm0.15$ (for $0.6<\log(m/\Ms)<2.0$) over the same projected area \citep[]{pang2013}.
Downwards the break, $\Gamma\approx-1.35$, the canonical value. This break,
which has been anticipated by \citet[]{pk2013}, can be attributed
to the switching of the binary period distribution to much tighter binaries for the massive stars
($m>5\Ms$; see Sec.~\ref{primbin}). The massive-stellar binaries being much more tightly bound
than the low-mass ones, the massive stars retain a much higher binary fraction with time
than the low-mass stars. Therefore, upwards the MF break, one excludes a good fraction of
lower-mass secondaries while detecting only the primaries, thereby
effectively flattening the PDMF, unlike downwards ($m<5\Ms$).
Fig.~\ref{fig:mfnow} (top and middle) demonstrates the
development of this break as an increasing number of low-mass binaries ionize.

Interestingly, the corresponding observed
PDMF shows a similar break \citep[]{pang2013} which is significant but less vivid and is at somewhat lower $m$.
In reality, it can be expected that such a
break in the PDMF would be moderately different than that obtained here. This is because
the change in the binary period distribution law would be more continuous than what is
assumed here. Note that \citet[]{pang2013} also obtained their (observed) $\Gamma$
upwards this break like the computed one here.

\section{Discussion and conclusion}\label{discuss}

This study demonstrates that it is possible to reproduce the detailed observed features of
the central cluster HD97950 of the NGC 3603 starburst region, within its age and mass constraints, from a
monolithic formation standpoint and adopting the reasonable and widely used values of the
parameters quantifying the gas dispersal phase \citep[]{pketl2001,sb2013}. It should be noted that the above results
imply a \emph{natural} match of the computed models with the observed data \emph{without any scaling or
fitting}.

The above calculations have been performed without applying an external Galactic potential.
As verified by comparing with calculations with a Solar-neighborhood-like field, the effect of
the external field is only minor. This can be expected as the highly compact initial system remains well
within its Roche lobe. Notably, the initial compact size ($r_h(0)\approx0.25$ pc) of the computed models
is consistent with the observed widths of the compact gas-filaments that host the proto-clusters \citep[]{andr2011}
and also with the mass-radius relation of embedded clusters \citep[]{mk2012}.

In this work, we assume the ``classical'' SFE of $\epsilon\approx33$\% over the proto-cluster
scale which is supported by both theoretical \citep[]{mnm2012} and observational \citep[]{lnl2003} studies.
We do not take into account the variations of SFE over the proto-star/subcluster scale as
suggested in recent theoretical studies \citep[]{giri2012,mok2012}. Such studies, based on hydrodynamic simulations,
indicate much higher local SFE which would influence the gas expulsion. However, inclusion of
proto-stellar outflows have been shown to substantially suppress the local SFE to $\lesssim 50$\% 
and is consistent with $\approx33$\% SFE over both local (stellar) and global (proto-cluster) scale
\citep[]{PriceBate2010,mnm2012}.

The analytic potential-lowering method applied here is widely used to this date for a good reason.
The self-consistent treatment of gas removal requires three-dimensional radiation-hydrodynamic calculations
which is formidable to date for the mass scale involved in this study.
High-resolution (reaching the ``opacity-limit'') smoothed-particle-hydrodynamics (SPH) computations
have been done so far forming stars in gas-spheres of only up to
$\approx500\Ms$ \citep[]{kl1998,bate2004,bate2009,giri2011,giri2012,bate2012}
but without any feedback and hence self-regulation mechanism. Radiation-magnetohydrodynamic (MHD) calculations
incorporating feedback to the star-forming gas has also been carried out from proto-stellar scales
\citep[]{mnm2012,bate2013} upto $\approx50\Ms$ gas spheres \citep[]{PriceBate2010}.
While the latter studies provide insights into the self-regulation mechanisms in the star formation process,
the processes that lead to the ultimate dispersal of the gas is still unclear.
Therefore, the analytic potential lowering is the best that can be done now for the mass scales involved here.

It should be noted, however, that the above monolithic cluster formation scenario
has nevertheless successfully reproduced ONC, Pleiades and Hyades clusters previously \citep[]{pketl2001}
and as well the R136 \citep[]{sb2013} and the HD97950 cluster as in here.
This indicates that although deprived of the details of star-formation and radiation-hydrodynamic processes
due to technological bottleneck, the adapted simplified gas-expulsion 
model and the choices of the parameters might still be an appropriate
description of the overall gas expulsion process. Indeed, in a comparative study, \citet[]{gb2001} found that analytic
gas-potential reduction has essentially the same effect on the cluster evolution as in an SPH computation
(for a given SFE) where the gas is removed via shock heating.
The present work thereby
provides another pillar to the to-date widely used classical gas expulsion scenario.

As such, the current state-of-the-art in theoretical research on star formation does not provide a definitive answer
to the gas dispersal and cluster formation processes and their interrelations. Given this, \emph{the present work
provides a plausible set of physical conditions that compellingly evolve to reproduce
the detailed observed properties of HD97950, subject to its photometric and age constraints.}
They comprise a \emph{solution} since the same computed model
reproduces multiple observed properties of the cluster
(it’s density profiles, velocity dispersion and stellar mass function).
Remarkably, the same conditions ($\epsilon\approx0.33$, $v_g\approx10{\rm~km}{~s}^{-1}$,
$\tau_g=r_h(0)/v_g$, $\tau_d\approx0.6$ Myr)
that reproduced the known properties of the ONC and
Pleiades \citep[]{pketl2001} also works for the much more massive R136 and HD97950 clusters
(see Table~\ref{tab1}).
The present work thereby serves as an intriguing evidence that these clusters might have formed
monolithically with the chosen parameters being an appropriate description of
the overall gas expulsion process.

Several authors alternatively propose VYMCs to be formed via hierarchical merging of
sub-clusters/structures \citep[]{fuji2012,sm2013,longm2014}.
In particular,
\citet[]{fuji2012} have demonstrated agreements with observed
radial cumulative distributions of massive stars and merger products in HD97950. This work,
although interesting, cannot be taken as a vivid counterexample because of notable drawbacks.
Here, the representative age of HD97950 has been taken to be 2.5 Myr
which is far too old
compared to the best-fitting age of $\approx 1$ Myr with only a small age spread, as obtained
from the cluster's PMS-MS CMD \citep[]{stol2004,pang2013}. In 1 Myr, the substructures
hardly merge to form a single cluster (\cf, Fig.~2 of \citealt[]{fuji2012}). Furthermore,
no attempt has been made to compare directly measurable quantities
like the structure and kinematics of the merged
cluster with observations. Hence,
it still remains to be seen how well a hierarchical merging scenario can reproduce the structure and
kinematics of HD97950 given its very young age. Based on the current theoretical literature, it is as such
impossible to rule out a single-starburst scenario of the formation of 
smooth-structured VYMCs in preference to a hierarchical one,
on fundamental grounds.

\begin{figure}
\includegraphics[width=9.2cm, height=6.3cm, angle=0]{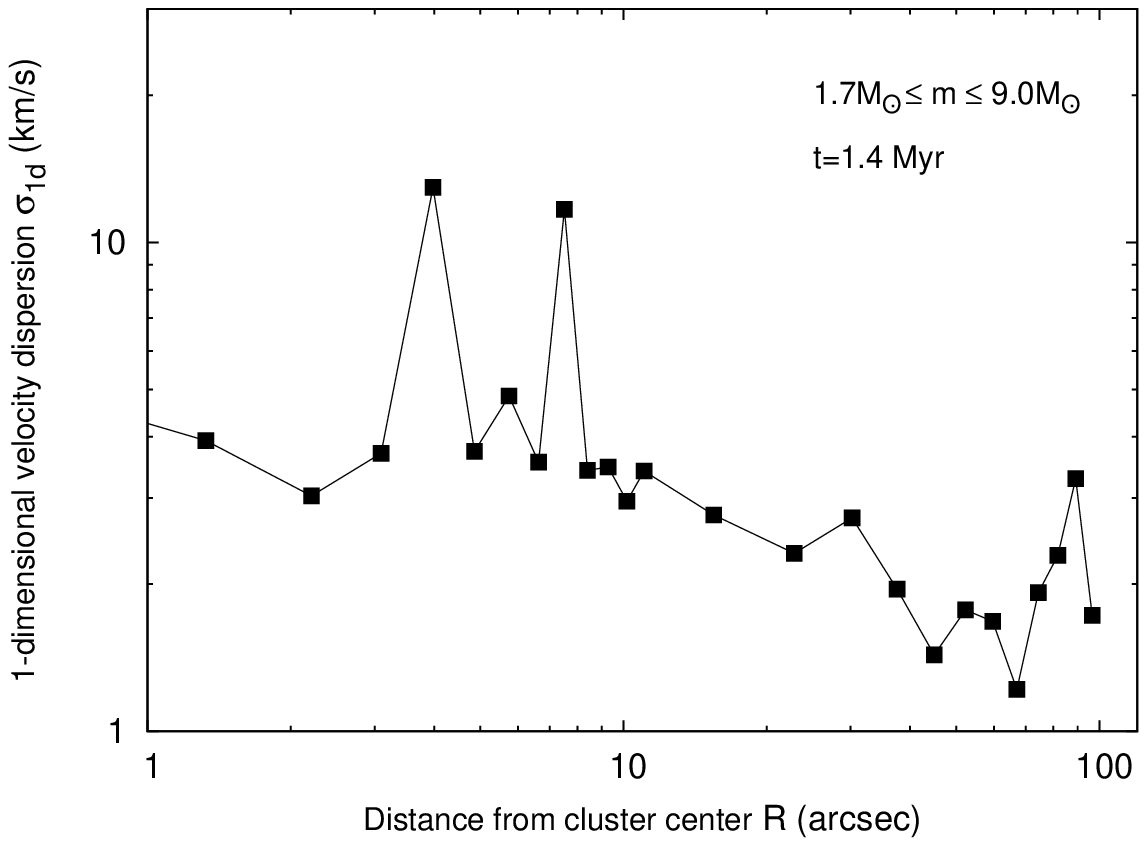}
\includegraphics[width=9.2cm, height=6.3cm, angle=0]{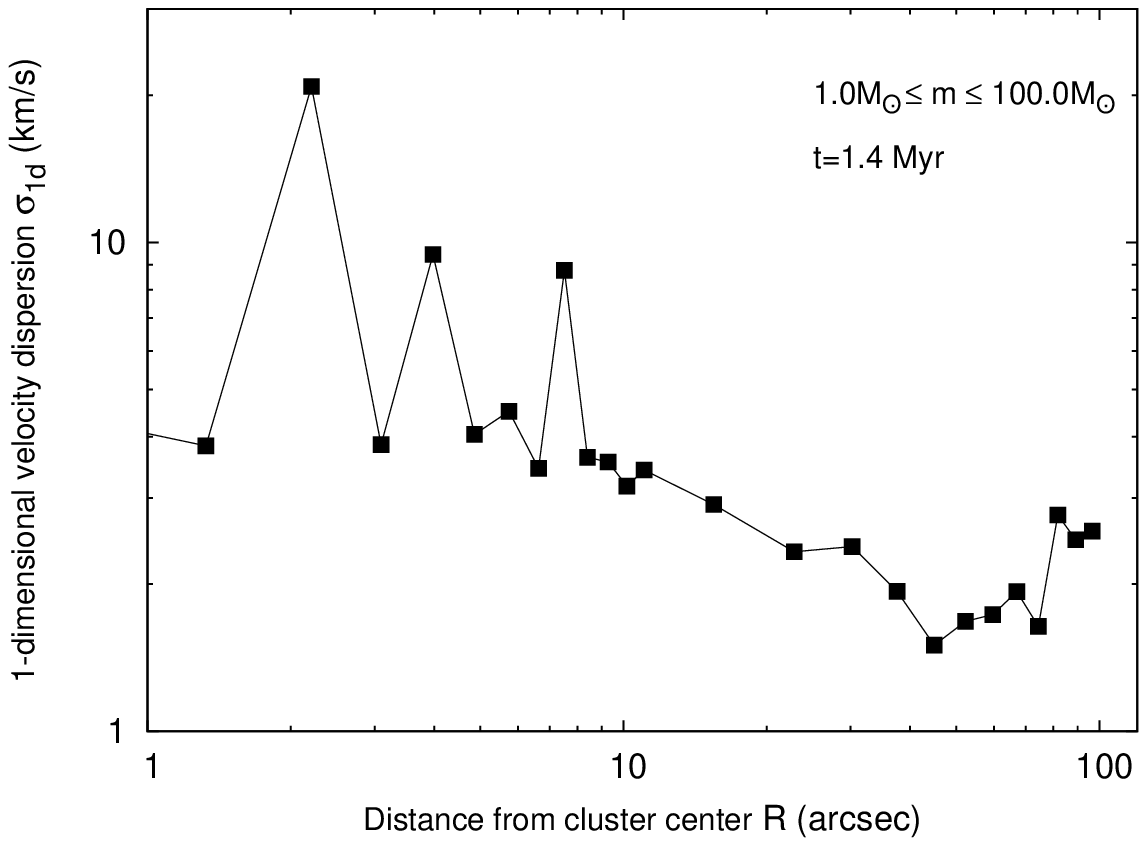}
\caption{Radial variation of 1-dimensional velocity dispersion, $\sigma_{\rm 1d}$, for the
computed model with primordial binaries (in presence of a tidal field) at $t=1.4$ Myr for stellar mass ranges
$1.7\Ms\leq m\leq9.0\Ms$ (top panel) and $1.0\Ms\leq m\leq100.0\Ms$ (bottom panel).
The overall increasing trend of $\sigma_{\rm 1d}$ with $R$ in the outer regions
($R\gtrsim40''$ in this case) indicates a recent gas expulsion. This can be tested in future
by more accurate proper motion measurements (\eg, by \emph{Gaia}) in the outer regions of the HD97950 cluster. 
The tangential velocities of selected stars in \citet[]{pang2013} (for $R\lesssim60''$)
do show an increasing trend with $R$. 
\label{fig:vprof}
}
\end{figure}

Notably, the construction of spatial density profiles involves accumulation over
finite-sized annuli and the heights over the profile depends on the choice of these annuli.
However, note that in Fig.~\ref{fig:Mdensprof}, the angular annuli are chosen nearly the same as those used
by \citet[]{hara2008} to obtain their observed profile. Also, in Fig.~\ref{fig:Ndensprof},
the same angular annuli are used to construct both of the number density profiles.

Several earlier authors have considered HD97950 to be in complete virial equilibrium and estimated
its dynamical mass to be $\approx20000\Ms$ from its observed velocity dispersion \citep[]{roch2010,pang2013}.
We have computed initially $M_{ecl}(0)\approx20000\Ms$ models (with $\epsilon\approx33$\%)
to find that such massive systems
yield either too centrally-dense or too expanded profiles, \ie, no well-matching initial conditions
could be found.
The $M_{ecl}(0)\approx10000\Ms$ models presented here, that do reproduce the observations,
are not in dynamical equilibrium at the epoch of matching ($t\approx1.4$ Myr), but are still expanding
(the innermost regions are close to re-collapsing). A prediction from this study, therefore, is
that the stars in the outer parts of HD97950 should have outgoing radial proper motions which can
be tested by future observations.

Fig.~\ref{fig:vprof} shows radial profiles of
$\sigma_{\rm 1d}$ at $t=1.4$ Myr for $1.7\Ms\leq m\leq9.0\Ms$ and $1.0\Ms\leq m\leq100.0\Ms$ where $\sigma_{\rm 1d}$
tends to increase for $R\gtrsim40''(\approx1.2{\rm~pc})$. The latter feature can be attributed to
the recent gas expulsion from the system resulting in its outer parts still expanding.  
Such a trend, which becomes more pronounced the closer the epoch of observation is to the gas expulsion, can
be tested by future, more accurate determinations of stellar proper motions in the outer regions of HD97950,
\eg, by \emph{Gaia}. Notably, the measured tangential velocities of selected stars in \citet[]{pang2013}
do show an increasing trend with radial distance in the outer region (measured up to $R\approx60''$).
The profiles in Fig.~\ref{fig:vprof} are obtained
from a computation with primordial binaries in presence of a solar-neighborhood-like external Galactic potential.

The above values of the parameters $\epsilon$, $\tau_d$ and $\tau_g$,
quantifying the overall gas expulsion, fair well with several young clusters (see above)
and hence can be considered well-representative. However, it would be of interest to consider
how our conclusions can be affected by moderate alterations of these parameters, representing possible
cluster-to-cluster variations. As shown in \citet[]{sb2013}, the effect of varying the delay time,
$\tau_d$, is simply a time-shift in the cluster's violent expansion without substantially affecting
the cluster's evolution for $t>\tau_d$. Hence, a delay of $\tau_d+\delta\tau_d$ in our computed models
would reproduce HD97950 at $\approx t+\delta\tau_d$. As such, one can allow for $\delta\tau_d\approx\pm0.1$ Myr
in the models computed here to obtain similarly good matchings. 

For $\epsilon<50$\%, as in the present case,
the bound fraction of stars that re-virializes to form the remnant cluster \citep[]{pketl2001,sb2013}
depends sensitively on $\tau_g$ \citep[]{bk2007}. However, the computed clusters here, which are still mostly expanding
at the HD97950's age of $t\approx1$ Myr, are far from re-gaining dynamical equilibrium (see above;
it takes $\approx3$ Myr to re-virialize an HD97950-like cluster as shown in \citealt[]{sb2013}) but
still retains a compact spatial distribution. Therefore, at $t\approx1$ Myr, most of the cluster's primordial
stellar population contributes to the computed radial profiles and the other quantities. Hence, the model-observation
agreements, as presented here, would still mostly persist with a moderate change of $\tau_g$ ($v_g$) as long as the
gas-removal is explosive (see Sec.~\ref{gdisp}). The computed values of $\sigma_{\rm 1d}$ can be expected to
increase moderately with decreasing $\tau_g$ but this trend is likely to be suppressed by the
substantial fluctuations in the values of $\sigma_{\rm 1d}$ (see Sec.~\ref{res}, Fig.~\ref{fig:v1d_pang}).

With a substantially higher $\epsilon$, one would expect a $M_{ecl}(0)\approx20000\Ms$
cluster to be even more overdense/oversized and a matching model would be of $M_{ecl}(0)\lesssim10000\Ms$, \ie,
below HD97950's photometric mass limit.
The reproducibility of the NGC3603's young cluster as a function of the parameters $\epsilon$, $\tau_d$ and $\tau_g$
will be investigated in detail in a future project.
An immediate improvement over this work would be to apply colour filters (\eg, V and I) to the
stellar luminosities to calculate the density profiles and the mass functions
for comparisons with observations even more accurately.

\acknowledgements

We thank Andera Stolte of the Argelander-Institut f\"ur Astronomie,
Bonn and Wolfgang Brandner of the Max-Planck-Institut f\"ur Astronomie, Heidelberg for
motivating discussions. We thank the anonymous referee for constructive suggestions
leading to substantial improvement of the presentation and the discussions.


\begin{thebibliography}{}

\bibitem[Aarseth(2003)]{ars2003}
Aarseth, S.J. 2003, ``Gravitational N-Body Simulations''. Cambridge University Press.

\bibitem[Adams(2000)]{adm2000}
Adams, F.C. 2000, \apj, 542, 964.

\bibitem[Anderson et al.(2009)]{andrson2009}
Anderson, L.D., Bania, T.M., Jackson, J.M., et al. 2009, \apjs, 181, 255.

\bibitem[Andr\'e et al.(2011)]{andr2011}
Andr\'e, P., Me\'nshchikov, A., Koenyves, V., et al. 2011,
in Alfaro Navarro, E.J., Gallego Calvente, A.T., Zapatero Osorio, M.R. (Eds.)
\emph{Stellar Clusters \& Associations: A RIA Workshop on Gaia}. 
Granada, Spain: IAA-CSIC, 321.

\bibitem[Banerjee \& Kroupa(2013)]{sb2013}
Banerjee, S. and Kroupa, P. 2013, \apj, 764, 29.

\bibitem[Bate \& Bonnell(2004)]{bate2004}
Bate M.R. and Bonnell, I.A. 2004, in Lamers, H.J.G.L.M., Smith, L.J., Nota A. (Eds.)
\emph{The Formation and Evolution of Massive Young Star Clusters},
(ASP Conf. Proc. 322). San Francisco: Astronomical Society of the Pacific, 289.

\bibitem[Bate(2009)]{bate2009}
Bate M.R., 2009, \mnras, 392, 590.

\bibitem[Bate(2012)]{bate2012}
Bate, M.R. 2012, \mnras, 419, 3115.

\bibitem[Bate et al.(2013)]{bate2013}
Bate, M.R., Tricco, T.S., Price, D.J. 2013, \mnras (doi:10.1093/mnras/stt1865).

\bibitem[\protect\citeauthoryear{Baumgardt \& Kroupa}{2007}]{bk2007}
Baumgardt, H. and Kroupa, P., 2007, \mnras, 380, 1589.

\bibitem[Boily \& Kroupa(2002)]{bokr2002}
Boily, C. and Kroupa, P. 2002, in
Grebel E.K., Brandner, W. (Eds.)
\emph{Modes of Star Formation and the Origin of Field Populations},
(ASP Conf. Proc. 285). San Francisco: Astronomical Society of the Pacific, 141. 

\bibitem[Chini et al.(2012)]{chini2012}
Chini, R., Hoffmeister, V.H., Nasseri, A., et al. 2012, \mnras, 424, 1925.

\bibitem[\protect\citeauthoryear{Churchwell}{2002}]{chrch2002}
Churchwell, E., 2002, \araa, 40, 27.

\bibitem[Fujii et al.(2012)]{fuji2012}
Fujii, M.S., Saitoh, T.R. and Portegies Zwart, S.F. 2012, \apj, 753, 85.

\bibitem[Geyer \& Burkert(2001)]{gb2001}
Geyer, M.P. and Burkert, A. 2001, \mnras, 323, 988.

\bibitem[Girichidis et al.(2011)]{giri2011}
Girichidis, P., Federrath, C., Banerjee, R. and Klessen, R.S. 2011, \mnras,  
413, 2741.

\bibitem[Girichidis et al.(2012)]{giri2012}
Girichidis, P., Federrath, C., Allison, R., et al. 2012, \mnras, 420, 3264.

\bibitem[Harayama et al.(2008)]{hara2008}
Harayama, Y., Eisenhauer, F. and Martins, F. 2008, \apj, 675, 1319.

\bibitem[Heggie \& Hut(2003)]{hh2003} 
Heggie, D.C. and Hut, P. 2003, ``The Gravitational Millon-Body Problem: A Multidisciplinary Approach to
Star Cluster Dynamics''. Cambridge University Press, Cambridge, UK.

\bibitem[Hennemann et al.(2012)]{henne2012}
Hennemann, M., Motte, F., Schneider, N., et al. 2012, \aap, 543, L3.

\bibitem[Hills(1980)]{hill80}
Hills, J.G. 1980, \apj, 235, 986.

\bibitem[Hill et al.(2011)]{hill2011} 
Hill, T., Motte, F., Didelon, P., et al. 2011 \aap, 533, A94.

\bibitem[Hurley et al.(2000)]{hur2000}
Hurley, J.R., Pols, O.R. and Tout, C.A. 2000, \mnras, 315, 543.

\bibitem[Klessen et al.(1998)]{kl1998}
Klessen, R.S., Burkert, A. and Bate, M.R. 1998, \apjl, 501, L205.

\bibitem[Kroupa(1995a)]{pk1995a}
Kroupa, P. 1995a, \mnras, 277, 1491.

\bibitem[Kroupa(1995b)]{pk1995b}
Kroupa, P. 1995b, \mnras, 277, 1507.

\bibitem[Kroupa et al.(1999)]{pk1999}
Kroupa, P., Petr, M.G. and McCaughrean M.J. 1999, NewA, 4, 495.

\bibitem[Kroupa(2001)]{pk2001}
Kroupa, P. 2001, \mnras, 322, 231.

\bibitem[Kroupa et al.(2001)]{pketl2001}
Kroupa, P. Aarseth, S. and Hurley, J. 2001, \mnras, 321, 699.

\bibitem[Kroupa(2005)]{pk2005}
Kroupa, P. 2005, in Turon, C., O'Flaherty, K.S. and Perryman, M.A.C. (Eds.)  
\emph{The Three-Dimensional Universe with Gaia} (ESA SP-576). 629p.

\bibitem[Kroupa(2008)]{pk2008}
Kroupa, P. 2008, in Aarseth, S.J., Mardling, R.A. and Tout, C.A. (Eds.)
\emph{The Cambridge N-body Lectures} (Lect. Notes Phys. 760). Springer-Verlag Berlin Heidelberg. 

\bibitem[Kroupa et al.(2013)]{pk2013}
Kroupa, P., Weidner, C., Pflamm-Altenburg, J., et al. 2013,
in Oswalt, T.D. and Gilmore, G. (Eds.)
\emph{Galactic Structure and Stellar Populations} (Planets, Stars and Stellar Systems, Volume 5).
Springer Science+Business Media Dordrecht (2013). 115p.

\bibitem[Krumholz \& Matzner(2009)]{krm2009}
Krumholz, M.R. and Matzner, C.D. 2009, \apj, 703, 1352.

\bibitem[K\"upper et al.(2011)]{kup2011}
K\"upper, A.H.W., Maschberger, T., Baumgardt, H. and Kroupa, P. 2011,
\mnras, 417, 2300.

\bibitem[Lada \& Lada(2003)]{lnl2003}
Lada, C.J. and  Lada, E.A. 2003, \araa, 41, 57.

\bibitem[Longmore et al.(2014)]{longm2014}
Longmore, S.N., Kruijssen, J.M.D., Bastian, N., et al. 2014,
in Beuther, H., Klessen, R., Dullemond, C. and Henning, Th. (Eds.)
\emph{Protostars and Planets VI}, University of Arizona Press, preprint (arXiv:1401.4175).

\bibitem[Machida \& Matsumoto(2012)]{mnm2012}
Machida, M.N. and Matsumoto, T. 2012, \mnras, 421, 588.

\bibitem[Malinen et al.(2012)]{mali2012}
Malinen, J., Juvela, M., Rawlings, M.G., et al. 2012, \aap, 544, A50.

\bibitem[Marks \& Kroupa(2011)]{mk2011}
Marks, M. and Kroupa, P. 2011, \mnras, 417, 1702.

\bibitem[Marks \& Kroupa(2012)]{mk2012}
Marks, M. and Kroupa, P. 2012, \aap, 543, A8.

\bibitem[Moeckel et al.(2012)]{mok2012}
Moeckel, N., Holland, C., Clarke, C.J. and Bonnell, I.A. 2012, \mnras, 425, 450.

\bibitem[Osterbrock(1965)]{ostb1965}
Osterbrock, D.E. 1965, \apj, 142, 1423.

\bibitem[Pang et al.(2013)]{pang2013}
Pang, X., Grebel, E.K., Allison, R., et al. 2013, \apj, 764, 73.

\bibitem[\protect\citeauthoryear{Patel et al.}{2005}]{pat2005}
Patel, N.A., Curiel, S., Sridharan, T.K., et al., 2005, \nat, 437, 109.

\bibitem[Pfalzner \& Kaczmarek(2013)]{pfkz2013}
Pfalzner, S. and Kaczmarek, T. 2013, \aap, 555, A135.

\bibitem[Portegies Zwart et al.(2010)]{pz2010}
Portegies Zwart, S.F., McMillan, S.L.W. and Gieles, M. 2010, \araa, 48, 431.

\bibitem[Price \& Bate(2010)]{PriceBate2010}
Price, D.J. and Bate, M.R. 2010, in
\emph{Plasmas in the laboratory and the Universe}:Interactions, Patterns, and Turbulence,
(AIP Conf. Proc. 1242), 205. 

\bibitem[Rochau et al.(2010)]{roch2010}
Rochau, B., Brandner, W., Stolte, A.,  Gennaro, M., et al. 2010, \apjl, 716, L90.

\bibitem[Sana \& Evans(2011)]{sev2011}
Sana, H. and Evans, C.J. 2011,
in Neiner, C., Wade, G., Meynet, G. and Peters, G. (Eds.)
\emph{Active OB Stars: Structure, Evolution, Mass Loss, and Critical Limits} (IAU Symp. 272).
Cambridge Univ. Press, Cambridge, 474.

\bibitem[Schneider et al.(2010)]{schn2010}
Schneider, N., Csengeri, T., Bontemps, S., et al. 2010, \aap, 520, A49.

\bibitem[Schneider et al.(2012)]{schn2012}
Schneider, N., Csengeri, T., Hennemann, M., et al. 2012, \aap, 540, L11.

\bibitem[Smith et al.(2013)]{sm2013}
Smith, R., Goodwin, S., Fellhauer, M. and Assmann, P. 2013, \mnras, 428, 1303.

\bibitem[Stolte et al.(2004)]{stol2004}
Stolte, A., Brandner, W., Brandl, B., et al. 2004, \aj, 128, 765.

\bibitem[Stolte et al.(2006)]{stol2006}
Stolte, A., Brandner, W., Brandl, B., et al. 2006, \aj, 132, 253.

\end{thebibliography}
\end{document}